\documentstyle[prd,aps,floats,epsf,fleqn]{revtex}

\newcommand{\npb}[2]{{\em Nucl. Phys.}              {\bf B#1}, #2 }

\newcommand{\prt}[2]{{\em Phys. Rev.}               {\bf D#1}, #2 }

\newcommand{\be}{\begin{equation}}
\newcommand{\ee}{\end{equation}}
\newcommand{\ba}{\begin{eqnarray}}
\newcommand{\ea}{\end{eqnarray}}

\def\al2{\frac{\alpha^2}{\pi^2}}

\begin{document}

\preprint{
\noindent
\hfill
\begin{minipage}[t]{3in}
\begin{flushright}
\vspace*{2cm}
\end{flushright}
\end{minipage}
}

\draft

\title{$S$-Matrix Elements in General Renormalization Schemes}
\author{Mingxing Luo}
\address{
Zhejiang Institute of Modern Physics, Department of Physics,
Zhejiang University, Hangzhou, Zhejiang 310027, P R China\\
Max Plank Institut fur Physik, Werner Heisenberg Institut, Theorie,
Fohringer Ring 6, D-80805 Munchen
}

\date{\today}

\maketitle

\begin{abstract}
Starting from the Lehmann-Symanzik-Zimmermann reduction theorem,
we provide a general procedure to extract $S$-matrix elements from Green functions
in arbitrary renormalization schemes.
\end{abstract}


In the framework of quantum field theories, renormalization is a necessity.
The bare parameters in the Lagrangian are infinite 
and cannot be used conveniently in physical predictions.
A re-parametrization of the theory in terms of finite variables is required.
To obtain finite Green functions, quantum fields themselves should be renormalized also.
Different re-parameterizations constitute different renormalization schemes.
For some physical processes, one scheme could be more convenient than others
though all schemes are equivalent.
It is hard to argue that there is a universally best scheme for all purposes.
To discuss different physical processes consistently,
such as in a global analysis of high precision electroweak experiments \cite{LLM},
all calculations need to be performed in one scheme.

One of the most frequently used schemes in the electroweak theory
is the so-called on-shell scheme \cite{sirlin},
in which physical masses of particles are used to parametrize the theory, 
coupling constants are defined in terms of certain scattering cross-sections at given energy scales,
and quantum fields are renormalized to give the two-point functions a residue of unity on the mass-poles. 
One big advantage of this scheme is that
$S$-matrix elements can be trivially obtained from the corresponding Green functions.
However, the Green functions are themselves complicated by 
the implementation of renormalization conditions.
More so, in theories such as softly-broken supersymmetric ones, 
the on-shell scheme cannot be realized for all
fields, due to over-constraints from symmetries.

It is thus expedient and sometimes necessary to introduce more general renormalization schemes.
In a general renormalization scheme,
the theory is parametrized by intermediate quantities that are not necessary physical observables
and renormalized fields are not required to give any particular value of residues on the mass-poles.
For example, in the modified minimal subtraction ($\overline{MS}$) scheme, 
one introduces the so-called $\overline{MS}$ parameters and renormalized fields,
which are obtained by subtracting the infinities and related $\log(4\pi)-\gamma$ terms
from the corresponding bare quantities \cite{msbar}.
In these schemes, Green functions assume simpler forms. 
However, care should be taken to obtain $S$-matrix elements.
In this note, 
we outline a general procedure to extract $S$-matrix elements from Green functions
in arbitrary renormalization schemes,
based upon the Lehmann-Symanzik-Zimmermann reduction theorem \cite{lsz}.

In some effective theories,
it could be convenient to keep non-canonical kinetic terms in the Lagrangian,
as the conversion to canonical forms may complicate other parts of the Lagrangian greatly.
Our procedure can be trivially extended to accommodate these cases.
On the other hand,
some calculations start by defining auxiliary quantities such as mixing angles between different fields.
These quantities are only well-defined at tree-level and not gauge invariant in general.
In our procedure, these quantities do not show up explicitly.
So they can be avoided in principle,
though it might be convenient for them to be introduced for phenomenological purposes.

To define $S$-matrix elements properly\cite{BD,Weinberg}, we separate the full Hamiltonian $H$ into two parts,
a free Hamiltonian $H_0$ and an interaction $H_{int}$,
$ 
H = H_0 + H_{int},
$ 
in such a way that $H$ and $H_0$ have the same eigenvalue spectrum.
For each eigenstate $\left|\Phi_\alpha^{(0)}\right\rangle$ of $H_0$ with eigenvalue $E_\alpha$,
one defines corresponding ``in" and ``out" states as eigenstates of $H$
\ba
H \left|\Phi_\alpha^{\pm}\right\rangle =  E_\alpha \left|\Phi_\alpha^{\pm}\right\rangle,
\ea
which satisfy the asymptotic condition
\ba
exp(-iH t) \int d\alpha g(\alpha) \left|\Phi_\alpha^{\pm}\right\rangle 
\rightarrow 
exp(-iH_0 t) \int d\alpha g(\alpha) \left|\Phi_\alpha^{(0)}\right\rangle, \nonumber
\ea
for $t \rightarrow - \infty$ and $t \rightarrow + \infty$, respectively.
Here $g(\alpha)$ is an arbitrary function but smoothly varying and non-zero over some finite range
$\Delta E$ of energy.
An $S$-matrix element is defined to be the transition probability amplitude from
an in-state $|\Phi_\alpha^{+}\rangle$ to an out-state $|\Phi_\beta^{-}\rangle$
\be
S_{\beta\alpha} = \left\langle \Phi_\beta^{-} \right.\left| \Phi_\alpha^{+} \right\rangle.
\ee

Following \cite{Weinberg}, we define an arbitrary Green function in momentum-space
\be
G(q_1 q_2 \cdots) =
\int_{FT}
\left\langle\Phi_0| T\left\{ A_1(x_1) A_2(x_2) \cdots A_n(x_n) \right\} |\Phi_0\right\rangle ,
\ee
where the $A$'s are Heisenberg-picture operators of arbitrary Lorentz type, $\Phi_0$ is the true vacuum, and
$\int_{FT}$ denotes integrations for Fourier transformations
\ba
\int_{FT} = \int d^4x_1  \cdots  d^4x_n e^{-iq_1 \cdot x_1}  \cdots  e^{-iq_n \cdot x_n}. \nonumber
\ea
If the $A$'s are ordinary fields appearing in the Lagrangian, then $G$ is a sum of terms calculated using
the ordinary Feynman rules, for all graphs with external lines corresponding to the fields $A$'s,
carrying off-shell four-momenta $q$'s into the graph.
We assume that the theory can be properly regularized and renormalized, so $G$ is well-defined.
It can be shown that $G$ has a pole at $\bar{s}=m^2-i \epsilon$, where $m$ is the mass of any one-particle state
$\Psi_{\vec{q}\sigma}$ that has non-vanishing matrix elements with states
$A_1^\dagger \Phi_0$ and $A_2 A_3 \cdots\Phi_0$, $\epsilon$ is a positive infinitesimal,
and the residue is given by\footnote
{Our normalization condition for one-particle states is
$
\sum_\sigma \int {d^3p \over (2\pi)^3} {1 \over \sqrt{2 p^0}} 
\left. |\Phi_{\vec{p},\sigma}\right\rangle \left\langle\Phi_{\vec{p},\sigma}| \right. = 1,
$
($p^0=\sqrt{\vec{p}^2+m^2}$),
instead of 
$
\sum_\sigma \int d^3p \left. |\Phi_{\vec{p},\sigma}\right\rangle \left\langle\Phi_{\vec{p},\sigma}| \right. = 1,
$
as that in \cite{Weinberg}.}
\be
G \rightarrow {i \over q^2 - \bar{s}} \sum_\sigma
\left\langle\Phi_0 | A_1(0) |\Phi_{\vec{q}\sigma}\right\rangle  
\int_{FT}
\left\langle\Phi_{\vec{q}\sigma}| T\left\{ A_2(x_2) \cdots \right\} |\Phi_0\right\rangle,
\label{pole-eq}
\ee
where the sum is over all spin states of the particle of mass $m$.
Depending upon the physics problem, $\Psi_{\vec{q}\sigma}$ can correspond to
either an in-state or an out-state.
If the particle is unstable and 
can decay into lighter particles of total decay width $\Gamma$,
the pole is displaced from the real axis by a finite amount, $\bar{s}=m^2- i m \Gamma$.

If $A_1$ has the Lorentz transformation properties of 
free field $\Psi_l$ belonging to an irreducible representation of the homogeneous Lorentz
group, as labeled by the subscript $l$, we can use Lorentz invariance to write
\be
\left\langle\Phi_0| A_1(0) |\Phi_{\vec{q}\sigma}\right\rangle  = N u_l(q,\sigma)
\ee
where $u_l(q,\sigma)$ is the coefficient function appearing in the free field $\psi_l$ 
and $N$ is a constant.
Define a truncated matrix element $M_l$ by
\be
\int_{FT} 
\left\langle\Phi_{\vec{q}\sigma}| T\left\{ A_2(x_2) \cdots \right\} |\Phi_0\right\rangle =
\sum_\sigma u^*_l(q,\sigma) M_l(q_2\cdots).
\ee
Then as $q^2 \rightarrow \bar{s}$
\be
G \rightarrow N {i \over q^2 - \bar{s} } \sum_{\sigma, m}
u_l(q,\sigma)  u^*_m(q,\sigma) M_m(q_2\cdots). \label{eqM}
\ee
The quantity multiplying $M_m$ in Eq. (\ref{eqM}) is the momentum space matrix propagator $\Delta_{l,m}(q)$
for the free field with the Lorentz transformation properties of $A_1$, so $M_{l}$ is the sum of all
graphs with external lines carrying momenta $q_1, q_2, \cdots$, corresponding to the operators
$A_1, A_2, \cdots$, but with the final propagator for the $A_1$ line stripped away.
Thus, to calculate the matrix for emission of a
particle from the sum of Feynman diagrams, one strips away the particle propagator and contracts
with the usual external line factor $u^*$.
This provides an alternative proof of the Lehmann-Symanzik-Zimmermann reduction theorem\cite{Weinberg},
which is applicable to cases of arbitrary spin.

If a theory has $N$ fields $\Psi^i_l$, which have the same Lorentz transformation properties
and other conserved quantum numbers as $\Psi_l$,
the two point functions between these fields are in general non-vanishing,
\be
(2\pi)^4 \delta^4(q_i+q_j) G^{ij}_{lm} (q_i) = 
\int_{FT}
\left\langle\Phi_0| T \left\{ \Psi^i_l(x_i) \Psi^{j\dagger}_m(x_j)  \right\} |\Phi_0\right\rangle.
\ee
Use the Lorentz invariance to write
\be
\left\langle\Phi_0| \Psi^i_l(0) |\Phi_{\vec{q}\sigma}\right\rangle  = N_i u_l(q,\sigma).
\ee
As $q^2 \rightarrow \bar{s}$, Eq. (\ref{pole-eq}) gives\footnote
{We assume that there is no degeneracy in mass for the same type of particles.
In case there is a degeneracy in mass, the particles will be distinguished by
their other quantum numbers and can be easily separated.
If there are particles with the same mass and other quantum numbers,
they can obviously be described by one quantum field.}
\be
G^{ij}_{lm} (q)
\rightarrow
N_i N_j^* {i \over q^2-\bar{s}} \sum_\sigma u_l(q,\sigma)  u^*_m(q,\sigma).
\label{Ndef}
\ee
\begin{figure}
\begin{center}
{\epsfxsize=2.500truein \epsfbox{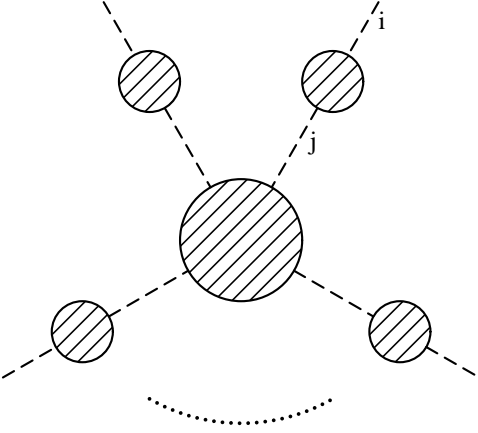}} 
\caption{\label{feynmanD} General structure of Feynman diagrams.}
\end{center}
\end{figure} 
On the other hand, an inspection of the structure of Feynman diagrams yields (Figure 1),
\be
G = \sum_{jm} G^{ij}_{lm} (q) \Gamma^j_m(q\cdots),
\ee
where $\Gamma^j_m$ is the sum of all Feynman diagrams
by stripping away the particle propagator $G^{ij}_{lm} (q)$.
As an intermediate step, we define a new truncated matrix element
\be
\sum_{jm} N_j^* u^*_m(q,\sigma) \Gamma^j_m(q\cdots).
\label{Sdef}
\ee
By repeating the process for all operators in $G$
and defining corresponding truncated matrix elements at each step, one finally gets the $S$-matrix element
up to an (irrelevant) overall phase factor.
This procedure has the salient feature which is independent of renormalization schemes.
In the on-shell scheme, one effectively defines a linear combination of $\Psi^i_l$ 
such that only one of the $N_i$'s is unity while all others vanish on each mass-pole.
Obviously this can only be performed recursively and is potentially tedious in practice.
Note that mixing angles between different fields
do not show up in Eq. (\ref{Sdef}) explicitly and can be avoided in principle.

We now express the $N$'s in terms of one-particle-irreducible (1PI) two-point functions.
To proceed, we need to be specific.
The 1PI two-point functions for scalar fields can be expressed as
\be
  \Gamma^B_{ij}(p^2) = p^2 Z_{ij}  - m^2_{ij} - \Sigma_{ij}(p^2),
\ee
where $Z_{ij}$ and $m^2_{ij}$ are coefficients of kinetic terms in the Lagrangian
and $\Sigma_{ij}$ are due to quantum loop contributions.
When normalized canonically, $Z$ is a unit matrix and $m^2$ is diagonal,
which are not required in our formalism.
Define $\gamma^B_{ij}$ as the residual matrix
of $\Gamma^B$ by crossing out its $j$-th row and the $i$-th column.
Denote $\Delta^B(p^2)=Det\left[\Gamma^B(p^2)\right]$
and $\Delta^B_{ij}(p^2)=(-)^{i+j}Det\left[\gamma^B_{ij}(p^2)\right]$.
The inverse of $\Gamma^B$ yields two-point functions of scalar fields up to a factor of $i$, 
\be
S_{ij}^B = {i \Delta^B_{ij}(p^2) \over \Delta^B(p^2)}.
\ee
$\bar{s}$ are determined from the equation $ \Delta^B(p^2) = 0$, its real part gives the mass-square of
the particle and the imaginary part the total decay width. The residue of $S_{ij}^B$ on the pole gives
\be
N_i N_j^* =  \lim_{p^2 \rightarrow \bar{s}} \left[ {d \Delta^B(p^2) \over dp^2} \right]^{-1}
 \Delta^B_{ij}(p^2),
\label{Nscalar}
\ee
from which the $N_i$'s are determined up to an (irrelevant) overall phase factor.
An application of Eqs. (\ref{Sdef}) and (\ref{Nscalar}) to the Higgs sector in the
Minimal Supersymmetric Standard Model readily yields 
results in the literature  \cite{susy}.

For vector fields, only gauge theories are known to be consistent.
Gauge fields are either massless or acquire masses via spontaneously symmetry breaking,
which can be realized in one way by introducing non-vanishing vacuum expectation values of scalars.
The 1PI two-point functions for the transverse components of vector fields are in general
\be
  \Gamma_{ij}^{\mu\nu}(p^2) = - g^{\mu\nu} \Gamma^V_{ij}(p^2) + \cdots,
\ee
from which the $N$'s can be obtained in the same manner as that of scalar fields.
For their longitudinal components $V^{i\mu}_L$, the situation is more subtle.
They have non-physical poles which are gauge dependent.
Fortunately, these non-physical poles are canceled exactly by
the same ones from the would-be Goldstone bosons due to BRST invariance \cite{BRST},
so they decouple from the $S$-matrix elements.
Accordingly, the Higgs boson masses can be selected from the whole set of scalar field poles
by excluding the would-be Goldstone bosons,
which are identical to the ones of $V^{i\mu}_L$'s.
However,
the correlation functions between $V^{i\mu}_L$'s and scalar fields need to be included
in calculating the $S$-matrix elements of Higgs bosons.

For Dirac fields without parity violation, 
one gets by Lorentz invariance,
\ba
\Gamma^D_{ij}(p) &=& \not{p} Z_{ij}  - m_{ij} - \not{p}  \Sigma^{(1)}_{ij}(p^2) - \Sigma^{(2)}_{ij}(p^2)
= \not{p} \Pi_{ij}(p^2)   - \Sigma_{ij}(p^2)
\ea
Define $\bar{\Gamma} = p^2  \Pi - \Sigma \Pi^{-1} \Sigma$ and its
residual matrix $\bar{\gamma}_{ij}$  by crossing out its $j$-th row and $i$-th column.
Denote $\bar{\Delta}(p^2)=Det\left[\bar{\Gamma}(p^2)\right]$
and $\bar{\Delta}_{ij}(p^2)=(-)^{i+j} Det\left[\bar{\gamma}_{ij}(p^2)\right]$.
The inverse of $\Gamma^D$ yields two-point functions of Dirac fields up to a factor of $i$, 
\be
S_{ij}^D = {i \over \bar{\Delta}} 
 \left[ \not{p} \bar{\Delta}_{ij} + (\Pi^{-1}\Sigma)_{ik} \bar{\Delta}_{kj} \right] 
\ee     
The pole position $\bar{s}$ is determined from the equation $\bar{\Delta}(p^2) = 0$
and the residue of $S_{ij}^D$ on the pole gives
\be
N_i N_j^* = \lim_{p^2 \rightarrow \bar{s}} \left[ {d \bar{\Delta} \over dp^2} \right]^{-1}
 \bar{\Delta}_{ij}.
\ee

For Dirac fields with parity violation, 
\ba
\Gamma^F_{ij}(p) & = &
 \left[ \not{p} Z^L_{ij}  - m^L_{ij} - \not{p}  \Sigma^{L(1)}_{ij}(p^2) - \Sigma^{L(2)}_{ij}(p^2)
   \right] P_L  
+\left[ \not{p} Z^R_{ij}  - m^R_{ij} - \not{p}  \Sigma^{R(1)}_{ij}(p^2) - \Sigma^{R(2)}_{ij}(p^2)
   \right] P_R \nonumber \\
&=& \left[ \not{p} \Pi_{ij}^L(p^2) -  \Sigma_{ij}^L(p^2) \right] P_L 
  + \left[ \not{p} \Pi_{ij}^R(p^2) -  \Sigma_{ij}^R(p^2) \right] P_R 
\ea
where $P_{L,R}=(1\mp\gamma_5)/2$ are the projection matrices.
Define $\Gamma^{R,L} = p^2  \Pi^{R,L} - \Sigma^{L,R} (\Pi^{L,R})^{-1} \Sigma^{R,L} $ and their
residual matrices $\gamma_{ij}^{R,L}$ by crossing out their $j$-th rows and $i$-th columns.
Denote $\Delta^{L,R}(p^2)=Det\left[\Gamma^{L,R}(p^2)\right]$
and $\Delta^{L,R}_{ij}(p^2)=(-)^{i+j} Det\left[\gamma^{L,R}_{ij}(p^2)\right]$.
The two-point functions are again the inverse of $\Gamma^F$ up to a factor $i$,
\ba
S^F_{ij} & = & {i \over \Delta^R} 
 \left\{ \not{p} \Delta^R_{ij} + \left[(\Pi^L)^{-1}\Sigma^R \right]_{ik} \Delta^R_{kj} \right\} P_L
+{i \over \Delta^L}
 \left\{ \not{p} \Delta^L_{ij} + \left[(\Pi^R)^{-1}\Sigma^L \right]_{ik} \Delta^L_{kj} \right\} P_R 
\ea
From $\Delta^R(p^2) = 0$ and $\Delta^L(p^2) = 0$, one gets the same set of pole positions.
From the correspondence of pole positions,
one identifies the left- and right- handed components of Dirac fermions.
Define $N_{iL,R}$
\be
\left\langle\Phi_0| \psi_{iL,R}(0) |\Phi_{\vec{q}L,R}\right\rangle  = N_{iL,R} u_{L,R}(q)
\ee  
The residue of $S_{ij}^F$ on the pole gives
\ba
N_{iL,R} N_{jL,R}^* &=& \lim_{p^2 \rightarrow \bar{s}} 
\left[  {d \Delta^{L,R}(p^2) \over dp^2} \right]^{-1} \Delta^{L,R}_{ij}(p^2),
\ea
from which we determine the $N$'s up to an overall phase factor 
for left- and right-handed fermions, respectively.  
These results also apply for Majorana fields, for which 
$\Pi^L = \left( \Pi^R \right)^*$ and $ \Sigma^L =\left( \Sigma^R \right)^*$, 
so  $N_{jL}=N_{jR}^*$. 

In summary, we have in Eqs. (\ref{Ndef}) and (\ref{Sdef}) provided a 
general procedure to extract $S$-matrix elements from Green functions
in arbitrary renormalization schemes. 
Furthermore, we have determined the normalization coefficients $N$
for scalar, vector, and various types of spin-1/2 fermion fields.
The analysis can be readily extend to cases of arbitrary spin
and our formalism is applicable to calculations in arbitrary quantum field theories.
It should be, in particular, useful for sophisticated theories such as
softly-broken supersymmetric ones,
where a full realization of the on-shell scheme is complicated if not impossible.

{\bf Acknowledgments:}
We would like to thank W.~Hollik for valuable discussions 
and to acknowledge the hospitality of Max Plank Institut fur Physik (Werner Heisenberg Institut).
This work was supported in part by CNSF-90103009
and a fund for Trans-Century Talents.


\end{document}